\documentclass[twoside,twocolumn]{article}
\usepackage[numbers]{natbib} % Citing by numbers
\usepackage{abstract} % Allows abstract customization
\usepackage{titlesec} % Allows customization of titles
\renewcommand\thesection{\Roman{section}} % Roman numerals for the sections
\renewcommand\thesubsection{\roman{subsection}} % roman numerals for subsections
\titleformat{\section}[block]{\large\scshape\centering}{\thesection.}{1em}{} % Change the look of the section titles
\titleformat{\subsection}[block]{\large}{\thesubsection.}{1em}{} % Change the look of the section titles

\usepackage{titling} % Customizing the title section

\usepackage[english]{babel}
\usepackage[utf8]{inputenc}
\usepackage[T1]{fontenc}
\usepackage[pdftex]{hyperref}

\usepackage{amssymb}
\usepackage{amsmath}
\usepackage{amsthm}
\usepackage{array}

\usepackage{multirow}
\usepackage{adjustbox} % vertical alignment of boxes (like minipages)
\usepackage{stfloats} % to place floats* at the bottom
\usepackage[caption=false]{subfig} % for subfloats
\usepackage{booktabs} % toprule, midrule and bottomrule
\usepackage{url}
\usepackage{wrapfig} % for figures inside text
\usepackage{pgfplots} % charts
\usepackage{tikz}
\usetikzlibrary{patterns} % patterns to fill elements
\usepgfplotslibrary{fillbetween} % fill areas

\usepackage[colorinlistoftodos]{todonotes} % TODOs or comments
\usepackage{xcolor} % to color the text
\usepackage{colortbl} % to color table cells
\usepackage{pifont} % to redefine cmark and xmark
\usepackage[misc,geometry]{ifsym} % for corresponding author envelope

\hypersetup{pdfborder=0 0 0}
\newcolumntype{L}[1]{>{\raggedright\arraybackslash}p{#1}}
\newcolumntype{C}[1]{>{\centering\arraybackslash}p{#1}}
\newcolumntype{R}[1]{>{\raggedleft\arraybackslash}p{#1}}
\newcommand{\xmark}{\ding{55}}%

\newcommand{\sectionname}{Sec.}
\definecolor{customblue}{HTML}{BDE0FF}
\definecolor{customgray}{HTML}{E6E6E6}
\definecolor{customred}{HTML}{FFCCCC}

\pgfplotsset{compat=1.13}

\makeatletter
\renewcommand{\@makefntext}[1]{%
  \setlength{\parindent}{0pt}%
  \begin{list}{}{%
    \setlength{\labelwidth}{1.5em}% <===================================
    \setlength{\leftmargin}{\labelwidth}%
    \setlength{\labelsep}{3pt}%
    \setlength{\itemsep}{0pt}%
    \setlength{\parsep}{0pt}%
    \setlength{\topsep}{0pt}%
    \footnotesize}%
  \item[\@makefnmark\hfil]#1%
  \end{list}%
}
\makeatother

\theoremstyle{definition}
\newtheorem{defn}{Definition}[]

\makeatletter
\def\th@definition{
  \thm@notefont{}% same as heading font
  \normalfont % body font
}
\makeatother

\begin{document}

    \graphicspath{{figures/}} % las figuras en la carpeta figures, ya no hace falta poner figures en los paths

    %!TEX root = main.tex

\title{Repairing Activity Start Times to Improve Business Process Simulation}

\author{
\textsc{David Chapela-Campa} \\[1ex]
\normalsize University of Tartu \\
\normalsize \href{mailto:david.chapela@ut.ee}{david.chapela@ut.ee}
\and
\textsc{Marlon Dumas} \\[1ex]
\normalsize University of Tartu \\
\normalsize \href{mailto:marlon.dumas@ut.ee}{marlon.dumas@ut.ee}
}

\date{}

\renewcommand{\maketitlehookd}{

\begin{abstract}

\noindent Business Process Simulation (BPS) is a common technique to estimate the impact of business process changes, e.g.\ what would be the cycle time of a process if the number of traces increases? %or if some resources become unavailable?
The starting point of BPS is a business process model annotated with simulation parameters (a BPS model).
%In traditional approaches, BPS models are manually designed by modeling specialists.
%This approach is time-consuming and error-prone. 
%To address this shortcoming, several studies have proposed methods to automatically discover BPS models from event logs -- extracted from enterprise information systems -- via process mining techniques.
Several studies have proposed methods to automatically discover BPS models from event logs -- extracted from enterprise information systems -- via process mining techniques.
These approaches model the processing time of each activity based on the start and end timestamps recorded in the event log.
In practice, however, it is common that the recorded start times do not precisely reflect the actual start of the activities.
For example, a resource starts working on an activity, but its start time is not recorded until she/he interacts with the system.
If not corrected, these situations induce waiting times in which the resource is considered to be free, while she/he is actually working.
To address this limitation, this article proposes a technique to identify the waiting time previous to each activity instance in which the resource is actually working on them, and repair their start time so that they reflect the actual processing time.
The idea of the proposed technique is that, as far as simulation is concerned, an activity instance may start once it is enabled and the corresponding resource is available.
Accordingly, for each activity instance, the proposed technique estimates the activity enablement and the resource availability time based on the information available in the event log, and repairs the start time to include the non-recorded processing time.
An empirical evaluation involving eight real-life event logs shows that the proposed approach leads to BPS models that closely reflect the temporal dynamics of the process.

%\keywords{business process simulation \and process mining \and start time estimation}

\end{abstract}

}
 % input instead of include to avoid auto \breakpage

    \maketitle

    %!TEX root = main.tex

\section{Introduction\label{sec:introduction}}

% General intro about Business Process Simulation

Business Process Simulation (BPS) is a technique for quantitative analysis of business processes~\cite{DBLP:books/sp/DumasRMR18}.
BPS is commonly used to estimate the impact of business process changes prior to their implementation~\cite{DBLP:conf/simpda/AguirrePA12,DBLP:conf/apbpm/ChoSY14,DBLP:conf/bpm/JohnsonDKFR18,DBLP:journals/is/MansRWG13,DBLP:conf/bpm/WynnDFHA07}.
Specifically, BPS enables managers and analysts to address ``what-if'' questions, such as ``what would be the cycle time of the process if the rate of arrival of new traces doubles?'' or ``what if 10\% of the workforce becomes unavailable for an extended time period?''.

%analysis, is done by simulating the behavior of the process with different configurations ---e.g.\ adding more resources--- and analyzing their impact in the performance measures.
%These simulated execution traces can then be used to compute performance measures of the process, such as cycle time or resource utilization.

The starting point of BPS is a business process model -- e.g.\ in the Business Process Model and Notation (BPMN)\footnote{\url{https://www.bpmn.org/}} -- enhanced with simulation parameters~\cite{DBLP:journals/bise/RosenthalTS21}, such as the distribution of processing times of each activity in the process (a BPS model).
The usefulness of BPS directly depends on the quality of the BPS model used as input~\cite{DBLP:conf/rcis/MeneghelloFLAT22}.
Specifically, one expects that a simulation of a BPS model leads to a collection of simulated traces (herein a \emph{simulated log}), which resembles the execution traces observed in the real executions of the process.
A BPS model unable to mimic the observed behavior would be unsuitable for analyzing the process in different scenarios.
Traditionally, BPS models were manually created by modeling experts.
This task is time-consuming and error-prone~\cite{DBLP:journals/kais/MarusterB09}.
To tackle this shortcoming, several studies~\cite{DBLP:journals/bise/MartinDC16,DBLP:journals/is/RozinatMSA09} have advocated for the use of process mining techniques to automatically discover BPS models from business process event logs recorded by (process-aware) information systems~\cite{DBLP:journals/dss/CamargoDG20,DBLP:journals/peerj-cs/CamargoDR21}.
In this context, an \emph{event log} is a set of events recording the execution of activity instances pertaining to a process.
Each event consists of an identifier of the process instance (trace) it belongs to, the label of the executed activity, the stage of the execution it records -- e.g.\ start or end --, the resource who performed the activity, and possibly some additional attributes.
Given this input, automated BPS model discovery methods learn a BPS model that, when simulated, leads to a simulated log that replicates the real process, as recorded in the original log.

% State the problem

Previous studies have highlighted that the applicability of process mining techniques is often hampered by data quality issues~\cite{DBLP:conf/cidm/BoseMA13,DBLP:conf/pakdd/AndrewsW17,DBLP:conf/bpm/FischerGADWR20,DBLP:journals/is/SuriadiAHW17,DBLP:journals/dss/SuriadiOAH15}.
One recurrent issue is the lack of timestamps, and particularly, the fact that event logs encountered in practice often do not have start timestamps, but only end timestamps~\cite{DBLP:journals/dss/SuriadiOAH15}.
In other scenarios, start timestamps may be present, but they do not reflect the actual time when the corresponding resource started actively working on the activity.
For example, an employee might start working on an activity by preparing the required materials, but the system does not record the activity start until the employee first interacts with the system.
This recurrent data quality issue has a direct impact on methods for automated BPS model discovery, since these methods rely on both start and end timestamps to compute the distribution of processing times of the activities in the process.
For example, if the start time does not precisely record when the resources started working on the activity instances, two scenarios can occur.
In the first scenario, the discovered BPS model simulates the sojourn time -- i.e.\ the amount of time since the activity instance is ready to be executed until it finishes -- only considering the (wrongly) recorded processing time.
However, this leads to unrealistically short cycle times, as the waiting time and non-recorded processing time are not being considered.
In the second scenario, the discovered BPS model includes waiting time previous to the processing of the activity instance, in which the resource is considered to be free, while she/he is already working on the activity.
Nevertheless, this can lead to a wrong allocation of the resources, as the resource was actually blocked while working in the preparation of the activity.

% Proposal

%In this setting, this paper addresses the following problem: given an event log $L$, compute an enhanced event log $L^{\prime}$ with adjusted start timestamps including the processing time, such that $L^{\prime}$ can be used to learn a BPS model that accurately replicates the temporal behavior recorded in $L$.
%In this setting, this paper addresses the following problem: given an event log $L$ wherein each event does not have a start timestamp, but do have trace identifier, activity, end timestamp, and resource attributes, compute an enhanced event log $L^{\prime}$ where the events contain both start and end timestamps, such that $L^{\prime}$ can be used to learn a BPS model that accurately replicates the temporal behavior recorded in $L$.
To tackle this problem, we propose to identify, for each activity instance in an event log, the portion of its waiting time that corresponds to non-recorded processing time -- i.e.\ the resource was actually working on the activity --, and repair the start time to include it as processing time.
To do this, our proposal combines the information of the enablement time of each activity instance with the availability information of its resource in order to compute the earliest starting point.
The reasoning behind this is that, as far as simulation is concerned, an activity instance may start once it is enabled and the corresponding resource is available.
Accordingly, we propose to consider as non-recorded processing time the interval from the earliest starting point until the recorded start of the activity instance.
%Our proposal estimates the adjusted start time of an activity instance as the earliest time point when both the activity was enabled, and the resource who performed it was available to start it.
%In this way, we include the exogenous waiting times in the processing time of the activity, allowing the simulation techniques to know its real duration.

The paper reports on an empirical evaluation involving eight real-life event logs containing both start and end timestamps. 
The evaluation aims at determining if a BPS model discovered from an event log with adjusted start times leads to simulated logs that closely replicate the temporal dynamics recorded in the original log, relative to a BPS model discovered directly from the original log.

% Paper outline

The remainder of the article is structured as follows.
\sectionname~\ref{sec:related-work} gives an overview of prior research related to start time estimation.
\sectionname~\ref{sec:approach} presents the proposed approach to estimate activity start times.
Finally, \sectionname~\ref{sec:evaluation} discusses the empirical evaluation, and \sectionname~\ref{sec:conclusions} draws conclusions and sketches future work directions.

    %!TEX root = main.tex

\section{Related Work\label{sec:related-work}}

% Timestamp repair

The importance of timestamp quality in event logs, as well as different techniques to repair timestamp-related problems, have been previously researched~\cite{DBLP:conf/caise/DixitSAWHBA18,DBLP:conf/bpm/ConfortiRHA20}.
Dixit et al.~\cite{DBLP:conf/caise/DixitSAWHBA18} study the order of events that are recorded at the same time --i.e.\ sharing timestamp value --, and propose a method to identify their correct order.
Addressing the same problem, Conforti et al.~\cite{DBLP:conf/bpm/ConfortiRHA20} propose a technique to repair the timestamp information of the events being recorded at the same instant.
Their proposal first detects their correct order, and then adjust their timestamp relying on a multimodal distribution of the duration of process activities.
Nevertheless, these techniques do not address the problem stated in \sectionname~\ref{sec:introduction} so far, as they do not focus on repairing the start times by including as processing time the waiting time in which the resource is actually working on the activity.

No previous research has been done to repair the start times of an event log by including, as processing time, the waiting time in which the resource is actually working in the next activity instance.
Nonetheless, some works have been presented focusing on the estimation of start times in event logs with missing data, which is related to the repair of the start times of an event log that we propose in this paper.
One of the first works was presented by Wombacher et al.~\cite{DBLP:conf/soca/WombacherIH11,DBLP:conf/sac/WombacherI13}.
The authors estimate the start time of the activity instances in an event log with only end times, in order to model the duration of the activities in the process.
They propose to first estimate, for each activity instance, the instant in which its resource became available as the end time of the previous activity instance processed by the same resource.
Then, they compute the instant at which the activity instance became available as the end time of the previous activity instance in the control flow.
Finally, they estimate the start time of each activity instance as the latest of their resource availability and enablement time (i.e.\ its earliest starting point).

% Senderovich queue mining

Senderovich et al. in~\cite{DBLP:conf/bpm/SenderovichLHGM15} develop on the enablement time heuristic presented by Wombacher et al.~\cite{DBLP:conf/soca/WombacherIH11,DBLP:conf/sac/WombacherI13} extending it to consider parallel relations using a given process model.
They focus on the discovery of resource queues in event logs with partial information.
When the information about the activity instances' enablement is missing in the event log, they propose to estimate it as the end time of the previous non-parallel activity instance.
% Baseline: use the resource availability to estimate the start time
Zabka et al.~\cite{custom/ZabkaBA21} have also used one of the heuristics presented by Wombacher et al.~\cite{DBLP:conf/soca/WombacherIH11,DBLP:conf/sac/WombacherI13} to estimate missing start times.
They aim to calculate the workforce of a process by estimating the duration of its executed activities.
To do this, they rely on the resource availability, and estimate the durations of an activity instance as the difference between its end time and the end time of the last event performed by the same resource.
Then, they consider the mode of the estimated durations of an activity as the \textit{typical processing time} of that activity.
To finally set the start times of each activity instance, they define a threshold w.r.t. the typical processing time.
If the calculated duration of an activity instance does not exceed the defined threshold, they use the estimated duration to calculate the start time w.r.t. the recorded end time.
If the estimated duration exceeds the threshold, they estimate the start time using the typical processing time of that activity.
% CAiSE start estimation with waiting times
Fracca et al.~\cite{DBLP:conf/caise/FraccaLAT22} also developed on the heuristics presented by Wombacher et al.~\cite{DBLP:conf/soca/WombacherIH11,DBLP:conf/sac/WombacherI13} to obtain the earliest estimation for the start time of each activity instance in an event log.
Then, taking the recorded end time of each activity instance as the latest estimation --i.e.\ zero duration --, they perform an optimization process (using business process simulation) to find the point in time between these two instants that better estimates the recorded start of the activity instances.

In this paper, we rely on similar heuristics as the presented in these techniques.
However, they specialize on their use to estimate the start times in event logs with only end times, and their evaluation (if any) is performed by comparing the estimation with the actual recorded start.
On the other hand, we propose to use the heuristics to identify the waiting time in which the resources are actually working on the activity, and increase the recorded processing time accordingly.

% Goncalves et al.: Estimate activity duration by filling the time in between to system-recorded events.

Finally, another work related to start time estimation was presented by Gonçalves et al.~\cite{DBLP:conf/ipmu/GoncalvesASD16}.
The authors work with event logs with two types of events: events that correspond to human work (i.e.\ performed by human resources), and events that correspond to a system (i.e.\ automatically registered).
They aim to estimate the duration of the activities performed by human resources, taking as anchors the system-recorded events.
For this, they group the human-recorded events in between two consecutive system-recorded events, and propose three frameworks to perform the estimation: \textit{i)} consider the human-recorded events as the end times of those activities and, for each one, set its start time as its previous event; \textit{ii)} consider the recorded events as the start times of those activities and, for each one, set its end time as its following event; and \textit{iii)} split the time between the two system-recorded events equally for all the human-performed activities inside it.
This proposal, similar to the previously commented ones, specializes on estimating the start times of activities performed by human resources, but it requires system-recorded events to use them as anchors.
In contrast, in the present paper, we do not require the existence of such events.

    %!TEX root = main.tex

\section{Approach\label{sec:approach}}

\begin{table*}[t]
    \centering
    {\footnotesize
    \caption{Running example: fraction of an activity instance log with 10 activity instances storing the trace, the executed activity, the start and end times, and the resource.}
    \label{tab:activity-instance-log-example}
    \begin{tabular}{ccccc}
        \textbf{trace ($\varphi$)} & \textbf{activity ($\alpha$)} & \textbf{start time ($\tau_{s}$)} & \textbf{end time ($\tau_{e}$)}   & \textbf{resource ($\rho$)} \\ \toprule \toprule
        \multicolumn{5}{c}{$\vdots$}                                                                                             \\
        23             & Register Order                  & 2021-03-07 12:59:21   & 2021-03-07 13:05:37   & Fry              \\
        24             & Register Order                  & 2021-03-07 13:06:53   & 2021-03-07 13:12:11   & Fry              \\
        23             & Prepare Package                 & 2021-03-07 13:11:07   & 2021-03-07 14:17:29   & Leela            \\
        23             & Prepare Invoice                 & 2021-03-07 13:15:21   & 2021-03-07 14:21:56   & Bender           \\
        24             & Prepare Invoice                 & 2021-03-07 14:23:12   & 2021-03-07 15:43:01   & Bender           \\
        23             & Deliver Package                 & 2021-03-07 14:30:00   & 2021-03-07 16:34:10   & Leela            \\
        25             & Prepare Package                 & 2021-03-08 10:02:32   & 2021-03-08 10:31:00   & Zoidberg         \\
        24             & Prepare Package                 & 2021-03-08 10:35:53   & 2021-03-08 11:11:05   & Zoidberg         \\
        24             & Deliver Package                 & 2021-03-08 11:15:07   & 2021-03-08 14:37:06   & Leela            \\
        25             & Prepare Invoice                 & 2021-03-08 14:40:23   & 2021-03-08 14:57:48   & Leela            \\
        \multicolumn{5}{c}{$\vdots$}                                                                                             \\ \bottomrule
    \end{tabular}
    }
\end{table*}

%This section first introduces basic notions of process mining related to the problem that is being addressed.
%Then, it describes the proposed approach to identify the waiting time that corresponds to non-recorded processing time -- i.e.\ the resource was actually working on the activity --, and repair the start time to include it as processing time.

We consider a business process that involves a set of \textit{activities} $A$.
We denote each of these activities with $\alpha$.
An \textit{activity instance} $\varepsilon = (\varphi, \alpha, \tau_{s}, \tau_{e}, \rho)$ denotes an execution of the activity $\alpha$, where $\varphi$ identifies the \textit{process trace} (i.e.\ the execution of the process) to which this event belongs to, $\tau_{s}$ and $\tau_{e}$ denote, respectively, the instants in time in which this activity instance started and ended, and $\rho$ identifies the \textit{resource} that performed the event.
Accordingly, we write $\varphi(\varepsilon_{i})$, $\alpha(\varepsilon_{i})$, $\tau_{s}(\varepsilon_{i})$, $\tau_{e}(\varepsilon_{i})$, and $\rho(\varepsilon_{i})$ to denote, respectively, the process trace, the activity, the start time, the end time, and the resource associated with the activity instance $\varepsilon_{i}$.
An \textit{activity instance log} $L$ is a collection of activity instances recording the information of the execution of a set of traces of a business process.
\tablename~\ref{tab:activity-instance-log-example} shown an example of 10 activity instances from an activity instance log.

It must be noted that usually the process execution information is recorded in event logs, which record the information of an activity instance as independent events, one representing its start, and another one representing its end.
Furthermore, there are cases in which event logs also store events recording other stages of the activity instances' lifecycle, e.g.\ schedule.
In this paper, we work with the concept of activity instance log to group the information of each activity instance -- e.g.\ its start and end events -- in a single element.
The transformation from a typical event log to its corresponding activity instance log is straightforward, and it has been already described in other research~\cite{DBLP:journals/is/MartinPM21}.
Nevertheless, we will use both terms: \textit{event log} and \textit{activity instance log}, to refer to the recorded process information used as starting point; and \textit{event} and \textit{activity instance}, to refer to the recording of an activity execution.

The starting point of our approach is an event log $L$.
We propose to identify the instant in time in which the resource started working on the activity by combining the availability of its resource and the enablement time of the executed activity (developing on the ideas presented by Wombacher et al.~\cite{DBLP:conf/sac/WombacherI13}).
The result is an enhanced event log $L^{\prime}$ with the repaired start times, such that the waiting time of each activity instance that corresponds to non-recorded processing time -- i.e.\ the resource was actually working on the activity --, is included as processing time.
%\figurename~\ref{fig:approach-overview} depicts an overview of our approach.

%\begin{figure}[t]
%    \centering
%    \includegraphics[width=0.98\columnwidth]{approach-overview.pdf}
%    \caption{Overview of the approach presented in this paper, which takes as input an event log without start times ($\tau_{s}$), and estimates them by combining the information of the enablement time and the resource availability of each activity instance. The result is the corresponding event log with start time information.}
%    \label{fig:approach-overview}
%\end{figure}

\subsection{Resource availability}

% One of the factors useful to infer when an activity instance has started to be processed is the availability of the resource that performed it.
We assume that when a resource ends an activity and becomes available, she/he will start working in the next one.
This reasoning is also used by Wombacher et al.~\cite{DBLP:conf/sac/WombacherI13}, and later by Fracca et al.~\cite{DBLP:conf/caise/FraccaLAT22}, to estimate the duration of the activities in event logs with no start times, and by Zabka et al.~\cite{custom/ZabkaBA21} to calculate activity durations in order to check if the workforce of a process has decreased or increased after the pandemic.
Obviously, we recognize that the resource can use some time to rest before starting a new activity, or after finishing one.
Nevertheless, this time is an inherent part of the resource performance, and should be taken into account as processing time.

\begin{figure}[t]
    \centering
    \includegraphics[width=0.98\columnwidth]{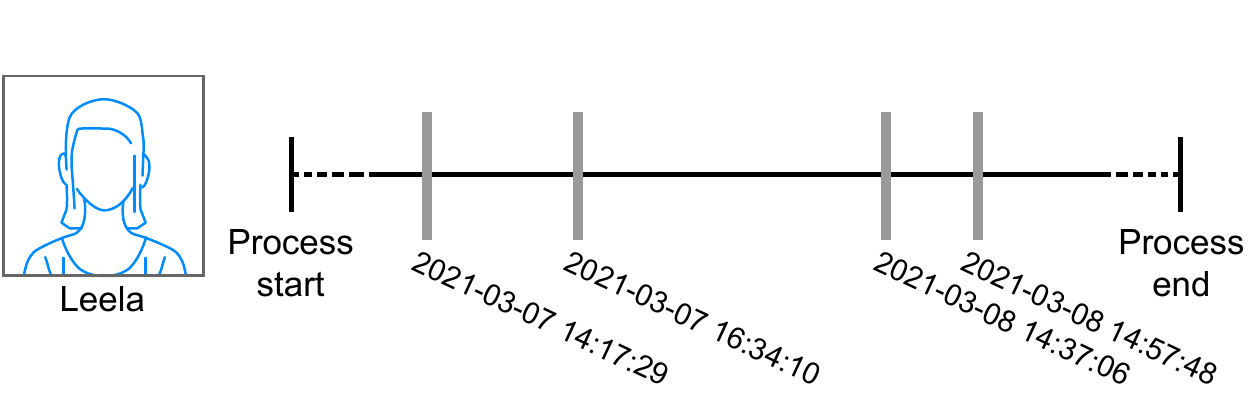}
    \caption{Individual timeline for the resource ``Leela'' corresponding to the example in \tablename~\ref{tab:activity-instance-log-example}, where each gray vertical line represents the end time of each activity instance performed by this resource.}
    \label{fig:resource-availability}
\end{figure}

To develop on this idea, we assume that the resources are dedicated only to the project under consideration, and that there is no multitasking (i.e.\ the resources only work in one activity at a time).
Otherwise, the information related to the resources' dedication to other projects, and their performance regarding multitasking, should be known.
%\footnote{\color{red}
%Given this assumption, we foresee that the proposed estimation technique will not be suitable for event logs where there is a high proportion of multitask activity instances.
%}
Following this reasoning, we define the resource availability time (\emph{rat}) of an activity instance as the earliest time in which its resource became available (i.e.\ the end time of the previous activity instance performed by the same resource).
To calculate this value, we build a timeline for each resource by adding all the registered end times of the activities she/he performed.
\figurename~\ref{fig:resource-availability} shows an example of this timeline for a resource in the activity instance log depicted in \tablename~\ref{tab:activity-instance-log-example}.
Accordingly, the resource availability time (\emph{rat}) of an activity instance is defined as follows:

\begin{defn}[Resource Availability Time\label{def:resource-availability-time}]

    Given an activity instance log $L$, the \textit{resource availability time} of an activity instance $\varepsilon = (\varphi, \alpha, \tau_{s}, \tau_{e}, \rho)$, such that $\varepsilon \in L$, is defined as $rat(\varepsilon) = max(\{ \tau_{e}(\varepsilon_{i}) \mid \varepsilon_{i} \in L \wedge \rho(\varepsilon_{i}) = \rho(\varepsilon) \wedge \tau_{e}(\varepsilon_{i}) < \tau_{e}(\varepsilon) \})$, i.e., the largest end time of those activity instances of $L$, processed by the same resource as $\varepsilon$, and being previous to its end.

\end{defn}

%It must be noted that the proposed method to calculate the resource availability assumes that after a resource finishes one activity, she/he becomes immediately available for processing the next one.
%Underpinning this assumption, there is the assumption that workers are only working on one activity at a time, in other words, there is no multitasking.
%Therefore, we foresee that the proposed estimation technique will not be suitable for event logs where there is a high proportion of multitask activity instances.

\subsection{Enablement time}

As explained before, we assume that when a resource ends an activity and becomes available, she/he will start working in the next one.
However, there might be cases in which the resource became available at one point, but could not start the next activity because other activities of the same process trace needed to finish before.
An example of this can be seen in \tablename~\ref{tab:activity-instance-log-example}, in which the activity \textit{Deliver Package} performed by \textit{Leela} on \textit{2021-03-08 14:37:06} could not start when she finished her previous activity (\textit{2021-03-07 16:34:10}), because the package of that delivery was not prepared until \textit{2021-03-08 11:11:05}.

To avoid this problem, and following with the reasoning of Wombacher et al.~\cite{DBLP:conf/sac/WombacherI13}, and later Fracca et al.~\cite{DBLP:conf/caise/FraccaLAT22}, we also consider the \textit{enablement time} of the executed activity.
The enablement time ($ent$) of an activity instance is the instant in time in which it became available for processing.
For example, in the event log from \tablename~\ref{tab:activity-instance-log-example}, an order has to be registered first, then the package and the invoice are prepared (in parallel) and, after that, the package can be delivered.
In this case, the delivery of the package cannot be performed until both the invoice and the package are prepared and, thus, its activity is not enabled for processing until that time.

A first approach to estimate the enablement time of an activity instance is to use the end time of the chronologically preceding activity instance (as proposed in~\cite{DBLP:conf/sac/WombacherI13,DBLP:conf/caise/FraccaLAT22}), assuming that the process follows a sequential order where each activity enables the following one.
Nevertheless, this observation does not hold if there are concurrency relations between the activities of the process -- a common situation in real-life processes.
A concurrency relation between two activities $A$ and $B$ means that these activities can be executed in any order, i.e. sometimes $A$ following $B$ and other times $B$ following $A$, thus no one enables the other\footnote{We note that other notions of concurrency have been proposed in the field of process mining.
An in-depth treatment of concurrency notions in process mining is provided in~\cite{DBLP:journals/toit/Armas-Cervantes19}. In this paper, we adopt a basic notion of interleaved concurrency between pairs of activities as defined above. The proposed approach is, however, modular and could be adapted to exploit more sophisticated notions of concurrency.}.
For example, in \tablename~\ref{tab:activity-instance-log-example}, activities \textit{Prepare Package} and \textit{Prepare invoice} are executed in any order (and sometimes even in parallel), as no one is required to start the other.

%Hence, in order to precisely measure the enablement time of an activity, we need to detect the concurrency relations between the activities of a process.
%A simple method to do this is to consider concurrent all activities $A$ and $B$ that appear, at least once, in both orders ($A$ following $B$, and $B$ following $A$).
%This concurrency detection method was proposed in one of the first discovery algorithms, the Alpha Miner~\cite{DBLP:journals/tkde/AalstWM04}.
%However, this method is too sensitive to outliers, as only one recording of $B$ following $A$, and another of $A$ following $B$, declares them as parallel even if one of them is observed 1 time, and the other 1,000.
%Furthermore, it also detects as concurrent the execution of length-2 loops such as $A$-$B$-$A$, as the two activities appear in both orders.
Senderovich et al.~\cite{DBLP:conf/bpm/SenderovichLHGM15} proposed to use a process model to identify the preceding activity(ies) of each activity, i.e.\ the activity(ies) enabling it.
However, this method requires a process model as input (of the discovery of one), and requires handling deviations from the process to identify the preceding activity when a trace does not follow a path represented in the model.
To remove this dependency, we propose to invert the analysis and use a concurrency oracle that indicate which activities are concurrent between them and, thus, not enabling each other.
In this way, the enablement time of an activity instance can be computed as the end time of the closest previous activity not being concurrent to it.
Accordingly, the enablement time of an activity instance can be defined as follows:

\begin{figure}[t]
    \centering
    \includegraphics[width=0.98\columnwidth]{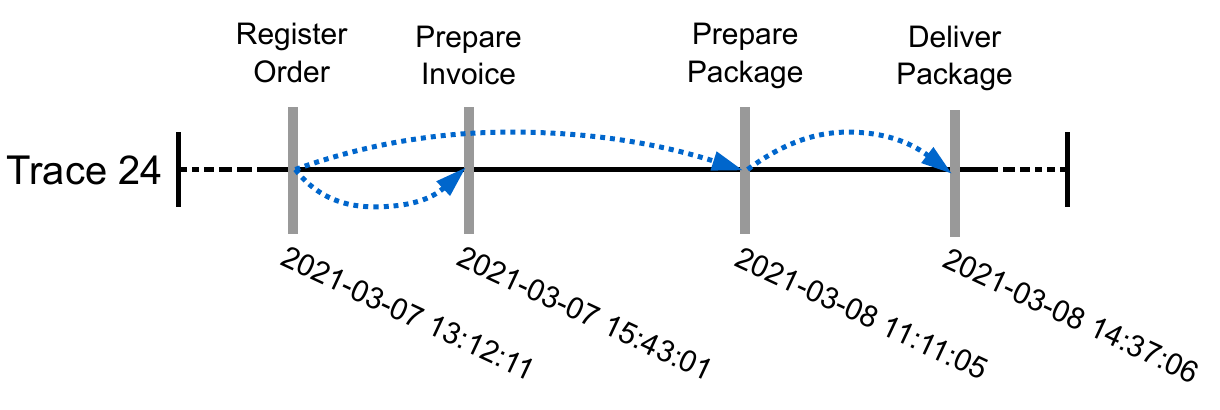}
    \caption{Timeline of trace 24 corresponding to the example in \tablename~\ref{tab:activity-instance-log-example}, where each gray vertical line represents the end time of each activity instance, and each dotted arrow connects an activity instance (source) with an activity instance it enables (target).}
    \label{fig:enablement-time}
\end{figure}

\begin{defn}[Enablement Time\label{def:enablement-time}]

    Given an activity instance log $L$, and a concurrency oracle that states if two activities $\alpha_{i}, \alpha_{j} \in L$ have ($\alpha_{i} \parallel \alpha_{j}$) or do not have ($\alpha_{i} \nparallel \alpha_{j}$) a concurrent relation, the \textit{enablement time} of an activity instance $\varepsilon = (\varphi, \alpha, \tau_{s}, \tau_{e}, \rho)$, such that $\varepsilon \in L$, is defined as $ent(\varepsilon) = max(\{ \tau_{e}(\varepsilon_{i}) \mid \varepsilon_{i} \in L \wedge \varphi(\varepsilon_{i}) = \varphi(\varepsilon) \wedge \tau_{e}(\varepsilon_{i}) < \tau_{e}(\varepsilon) \wedge \alpha(\varepsilon_{i}) \nparallel \alpha(\varepsilon) \})$, i.e., the largest end time of those activity instances of $L$, belonging to the same process trace, being previous to $\varepsilon$'s end time, and which activity does not have a concurrent relation with $\varepsilon$'s activity.
    
\end{defn}

We propose to use the concurrency oracle of the Heuristics Miner~\cite{DBLP:conf/cidm/WeijtersR11} algorithm to discover the concurrency relations between the process' activities.
This method computes a degree of confidence for each observed relation between two activities that it is a concurrent or directly-follows relation, based on the percentage of occurrences in each order.
Then, based on a set of defined thresholds, it retrieves the concurrent relations of the process.
It must be noted that our proposal to compute the enablement time is not dependent on the concurrency oracle implementation.
The concurrency relations could be alternatively extracted with other concurrency discovery algorithm, or from a process model, if available.

\figurename~\ref{fig:enablement-time} shows the timeline for one process trace, depicting the enablement relations between the activities.
As can be seen, both \textit{Prepare Invoice} and \textit{Prepare Package}, which have a concurrency relation between them, are enabled by \textit{Register Order}.
Finally, \textit{Deliver Package} is enabled by \textit{Prepare Package}, its previous activity not being concurrent to it.

\subsection{Start time repair}

\begin{figure}[t]
    \centering
    \includegraphics[width=0.98\columnwidth]{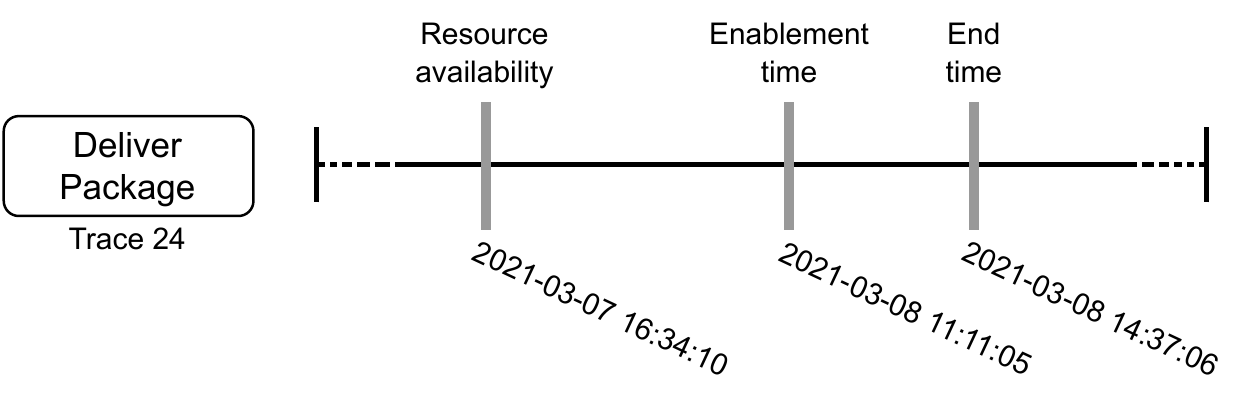}
    
    \vspace{10pt}
    
    \includegraphics[width=0.98\columnwidth]{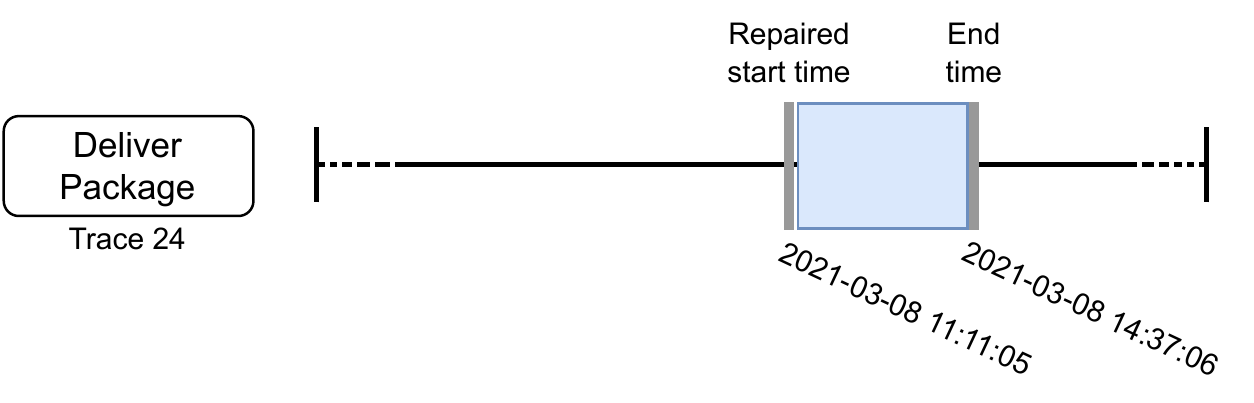}
    
    \caption{Timeline denoting the end time, resource availability, and enablement time for the activity instance ``Deliver Package'', of the process trace 24, corresponding to the example in \tablename~\ref{tab:activity-instance-log-example}.
    As well as its duration based on the estimated start time.}
    \label{fig:start-time-estimation}
\end{figure}

In order to include the non-recorded processing time into each activity instance processing time, we propose to set its start time to the earliest point in time in which the activity instance is enabled, and its resource is available for processing.
\figurename~\ref{fig:start-time-estimation} shows an example of this adjustment for an activity instance of the event log in \tablename~\ref{tab:activity-instance-log-example}.
Accordingly, the repair of an activity instance's start time can be defined as follows:

\begin{defn}[Start Time Repair\label{def:start-time}]

    Given an activity instance log $L$, the \textit{repaired start time} of an activity instance $\varepsilon = (\varphi, \alpha, \tau_{s}, \tau_{e}, \rho)$, such that $\varepsilon \in L$, is defined as $\tau^{\prime}_{s}(\varepsilon) = max(rat(\varepsilon), ent(\varepsilon))$, i.e., the largest of its enablement time and its resource availability time.

\end{defn}

There may exist cases in which an activity instance does not have a resource assigned.
In those cases, our technique assigns to the activity instance a resource pool with maximum capacity, meaning this that the activity can be executed as soon as it is enabled.
Thus, its repaired start time will be estimated as its enablement time.

In order to exclude the case of automatic activities which are executed at the same time as others, e.g.\ control automatic activities performed after the end of a subprocess, we offer the possibility to the user to indicate those resources that are "bot resources" and execute the activities instantly.
In the same way, we offer the possibility to identify activities that are always executed automatically.
Accordingly, we let the start time of those cases as their end time.

Developing on the idea presented by Zabka et al.~\cite{custom/ZabkaBA21}, we offer a threshold to control outlier estimations.
For this, we first estimate the earliest starting point of all activity instances.
Then, we calculate their most typical repaired duration (from the earliest starting point to the recorded end time).
Finally, given a threshold $\eta$, we establish an upper bound such that all activity instances with a repaired duration over $\eta$ times their typical repaired duration gets their earliest starting point re-estimated, so that their repaired duration is $\eta$ times their typical repaired duration.

    %!TEX root = main.tex

\section{Evaluation\label{sec:evaluation}}

This section reports on an experimental evaluation of the proposed approach to evaluate the impact of the proposed start time repair in the simulation results.

Our hypothesis is that our repaired start times will lead to simulated logs that replicate the temporal dynamics recorded in the original log better than a BPS model discovered directly from the original log.
The reasoning behind this is that the simulation does not take into account exogenous waiting times, e.g.\ waiting for the client to complete an activity or answer a question.
Thus, with the premise of an activity starting once the resource is available and the activity enabled, we include that information in the processing times (as an indirect temporal cost for the activities of the process), and the temporal similarity is expected to be better.
In this way, our technique generates user utilization that are reflective of the availability of the resource to perform the activities, calculating better the tempo-dynamics of the process.

It must be noted that our hypothesis of an activity starting when the resource is available works better under the assumption that the resources are internal, even better with resources dedicated only to the project under study.
If the resource is external ---e.g.\ the client, or a consultant from another company---, there is a possibility that an activity instance did not start even when the activity was enabled and the resource available, because he/she was busy with activities from other projects.
However, to prevent this from happening, the information of the resources' participation in other projects would have to be present in the event log.

% Note that, the objective of addressing these two research questions is not only to estimate the start times in event logs without them, allowing their simulation, but to offer a method to re-estimate start times in event logs with already recorded start times, improving the simulation results.

\subsection{Datasets}

In order to enhance the ecological validity of the empirical evaluation, we selected a set of real-life event logs with balanced complexity characteristics:

\begin{table*}[t]
    \centering
    {\footnotesize
    \caption{Characteristics of the event logs used in the evaluation: 4 real-life logs of low complexity and 4 real-life logs of high complexity. The multitasking column denotes the percentage of the resources' working time in which they are working in more than one activity.}
    \label{tab:log-characteristics}
    
    \begin{tabular}{l r r r r r r r r}
                & Type      & Complexity  & Traces   & \begin{tabular}[c]{@{}r@{}}Activity\\ instances\end{tabular} & Variants & Activities & Resources & Multitasking \\ \toprule \toprule
        ACR     & REAL      & Low         & 954      & 4,962                 & 99          & 16            & 559       & 20\%   \\
        BNK     & REAL      & Low         & 70,512   & 415,261               & 198         & 8             & 9         & 99\%   \\
        MP      & REAL      & Low         & 225      & 4,503                 & 217         & 24            & 46        & 20\%   \\
        CALL    & REAL      & Low         & 3,885    & 7,548                 & 354         & 6             & 48        & 52\%   \\ \midrule
        GOV     & REAL      & High        & 23,506   & 110,659               & 1,003       & 131           & 30        & 95\%   \\
        INS     & REAL      & High        & 1,182    & 23,141                & 1,097       & 9             & 136       & 49\%   \\
        BPIC12W & REAL      & High        & 8,616    & 59,302                & 2,115       & 6             & 57        & 9\%   \\
        BPIC17W & REAL      & High        & 30,276   & 240,854               & 6,356       & 8             & 148       & 3\%   \\ \bottomrule
    \end{tabular}
    }
\end{table*}

\begin{itemize} 
    \item Four real-life event logs of low complexity, with a number of variants ranging from 99 to 354.
    These logs are from an academic credentials' management process (ACR), a bank process (BNK), an industrial production line (MP), and a call center process (CALL).
    
    \item Four real-life event logs of high complexity, with a number of variants ranging from 1,000 to more than 6,000.
    One of these event logs records the execution of an approval application system of a governmental agency (GOV).
    Another one corresponds to an insurance process (INS).
    We also used the event log from the Business Process Intelligence Challenge of 2012~\cite{vanDongen_BPIC12}, recording a loan application process from a Dutch financial institution.
    We preprocessed this event log retaining only the events corresponding to activities performed by human resources (i.e.\ only activity instances that have a duration).
    Finally, we also used the updated version of the BPIC 2012 event log, of the Business Process Intelligence Challenge of 2017~\cite{vanDongen_BPIC17}.
    We preprocessed this event log by following the recommendations reported by the winning teams participating in the BPIC 2017 (\url{https://www.win.tue.nl/bpi/doku.php?id=2017:challenge}).
\end{itemize}

The main characteristics of the event logs used in this experimentation can be seen in \tablename~\ref{tab:log-characteristics}.
%The event logs of ACR, MP, BPIC12W, and BPIC17W datasets, the simulated event logs and their corresponding simulation models, and the experimentation results are available in [\textit{anonymized link}].
% \url{put-zenodo-utl-here}

\begin{table*}[t]
    \centering
    {\footnotesize
    \caption{Different configurations of the technique proposed in this paper, where $\overline{x}$ denotes the statistic used to calculate the most typical duration, and $\eta$ the outlier threshold.}
    \label{tab:configuration-techniques}
    \begin{tabular}{rccccccccc}
    \toprule                                                       \\
                   & \textbf{MED} & \textbf{MED-5} & \textbf{MED-2} & \textbf{MOD} & \textbf{MOD-5} & \textbf{MOD-2} \\ \toprule \toprule
    $\overline{x}$ & Median        & Median         & Median         & Mode          & Mode           & Mode           \\
    $\eta$         & \xmark        & 500\%          & 200\%          & \xmark        & 500\%          & 200\%          \\ \bottomrule
    \end{tabular}
    }
\end{table*}

\subsection{Experimental setup}

To evaluate the impact of outlier estimations -- e.g.\ due to the resource not working for a long period --, we used three different configurations of our proposal: \textit{i)} with no outlier threshold, meaning that the repaired start times are the final ones; \textit{ii)} with an outlier threshold of 2 (low threshold), meaning that all activity instances with a repaired duration exceeding 2 times their most typical repaired duration get their start time re-repaired to that duration; and \textit{iii)} with an outlier threshold of 5 (high threshold), so the re-repaired activity instances are those with a repaired duration over 5 times their most typical repaired duration.
For each configuration, we used both the median and the mode to compute the most typical repaired duration, in order to check if the results depend on the chosen statistics.
\tablename~\ref{tab:configuration-techniques} shows the configurations of all the evaluated techniques.

For each dataset, we run each of the configuration to repair the start times of each event log, obtaining six different repaired event logs (one per configuration).
Then, we run SIMOD~\cite{DBLP:journals/dss/CamargoDG20} to automatically discover a simulation process with each of the six repaired event logs, as well as with the original event log.
With this simulation process, we simulated five event logs with the same number of traces than the original one.
Finally, we measured the distance of each simulated log w.r.t. the original one (for each proposal, the result is the median of its 5 simulations).

To validate if the repaired start times improve the simulation results, we compared the distance of the simulated logs w.r.t. the original one -- the lower the distance is, the more similar the simulated log is w.r.t. the original log.
In previous work, such comparisons are made along two dimensions: the control-flow and the temporal dimensions~\cite{DBLP:journals/dss/CamargoDG20,DBLP:journals/peerj-cs/CamargoDR21}.
Given that a start time repair does not alter the control-flow of the process, we hereby focus on the temporal dimension.
Specifically, we compared each simulated log against the corresponding original log in two ways: \textit{i)} by comparing the distributions of their timestamps (both start and end timestamps); and \textit{ii)} by comparing the distribution of cycle times of their traces.\footnote{The cycle time of a trace (a.k.a. the trace duration) is the difference between the largest end timestamp and the smallest start timestamp observed in the trace.} 
To compare two distributions, we use the Wasserstein Distance -- Earth Movers' Distance (EMD) --, which measures the amount of movements that have to be applied to one discretized distribution to obtain the other, penalizing the movements by their distance. 

The EMD is defined over a pair of discretized distributions (i.e. histograms).
To discretize the distribution of timestamps, we extract all the start and end timestamps in the log, and we group them by date-hour.
For example, all timestamps that fall between 9 February 2022 at 10:00:00 and the same day at 10:59:59 are placed in one group (bin).
On the other hand, the cycle times are discretized as follows.
First, the cycle times of the original log are grouped into 100 equidistant bins.
Let $W$ be the width between these bins.
Each simulated log derived from this original log is then discretized into bins of width $W$.
Having discretized the timestamps and the cycle times of the original log and its simulated counterparts, we then calculate their EMD and use this measure to compare the simulation performed with the original log, with each of the six variants of the proposed technique.
Importantly, the magnitude of the EMD depends on the dataset (e.g. larger datasets will naturally give rise to larger EMD values).
Accordingly, we use the EMD to make intra-dataset comparisons only (i.e.\ to measure the relative performance of multiple techniques within the same dataset).

%and not to make inter-dataset comparisons (e.g.\ to compare the performance 

It must be noted that the similarity of the simulated logs using the original event log have an advantage, as their BPS models have been discovered by training with the event logs then used for the similarity evaluation.
On the other hand, the BPS models of the other simulated logs have been trained with the repaired event logs.

\subsection{Results}

\begin{table*}[t]
    \centering
    {\footnotesize
    \caption{EMD of the timestamps for the simulated logs w.r.t. the original log (the lower, the more similar the simulated logs are w.r.t. the original log). The gray cells mark, for each dataset, the best result.}
    \label{tab:results-emd}
    \begin{tabular}{lrrrrrrrr}
               & \textbf{ACR}                & \textbf{BNK}                & \textbf{MP}                & \textbf{CALL}              & \textbf{GOV}                 & \textbf{INS}                  & \textbf{BPIC12W}            & \textbf{BPIC17W}              \\ \toprule \toprule
    ORIG       & 456.32                      & 871.77                      & \cellcolor{gray!45}94.16   & 1,338.05                   & \cellcolor{gray!45}2,519.65   & 9,466.29                      & 2,170.63                    & 5,823.61                      \\
    MED        & 2,793.60                    & 9,476.72                    & 4,703.49                   & 80.77                      & 6,658.81                      & \cellcolor{gray!45}3,709.33   & 3,835.93                    & 925.40                        \\
    MED-5      & 2,770.54                    & \cellcolor{gray!45}408.45   & 1,988.31                   & \cellcolor{gray!45}73.08   & 6,330.92                      & 9,242.26                      & \cellcolor{gray!45}166.80   & 5,830.43                      \\
    MED-2      & 2,821.47                    & 1,897.83                    & 964.11                     & 112.83                     & 6,574.81                      & 9,182.91                      & 2,558.82                    & 2,085.34                      \\
    MOD        & \cellcolor{gray!45}130.02   & 3,108.90                    & 2,238.73                   & 93.88                      & 6,656.59                      & 8,495.60                      & 686.85                      & \cellcolor{gray!45}789.32     \\
    MOD-5      & 372.14                      & 2,672.34                    & 2,276.79                   & 75.06                      & 6,644.68                      & 8,801.83                      & 2,383.89                    & 2,094.49                      \\
    MOD-2      & 2,989.33                    & 1,612.38                    & 1,529.58                   & 112.90                     & 6,686.00                      & 8,362.68                      & 2,332.21                    & 2,247.24                      \\ \bottomrule
    \end{tabular}
    }
\end{table*}

\begin{figure*}[b]
    \begin{tikzpicture}
        \begin{axis}[
            ybar,
            height = 6cm,
            width = 17cm,
            bar width=0.1pt,
            ylabel = {Number of events},
            ylabel style= {yshift=-1mm},
            ymin = 0,
            xlabel = {Timeline},
            xlabel style= {yshift=1mm},
            xmin = 0,
            xticklabels = \empty,
            legend entries = {ORIG,S-ORIG,S-MODE}
        ]
            \addplot +[
                blue!40,
                fill opacity=0.4
            ] table[] {data/bpic17-start-end_ORIG.dat};
            \addplot +[
                red!60,
                fill opacity=0.4
            ] table[] {data/bpic17-start-end_SORIG.dat};
            \addplot +[
                green!30,
                fill opacity=0.4
            ] table[] {data/bpic17-start-end_SMODE.dat};
            \legend{ORIG,S-ORIG,S-MODE}
        \end{axis}
    \end{tikzpicture}

    \caption{Histograms of the start-end timestamps for the BPIC17W dataset.
    ORIG corresponds to the timestamps of the original event log.
    S-ORIG corresponds to the timestamps of the event log simulated with the original one.
    S-MODE corresponds to the timestamps of the event log simulated with the repaired event log (no outlier threshold and using the mode as the statistic).}
    \label{fig:chart-start-end-distribution}
\end{figure*}
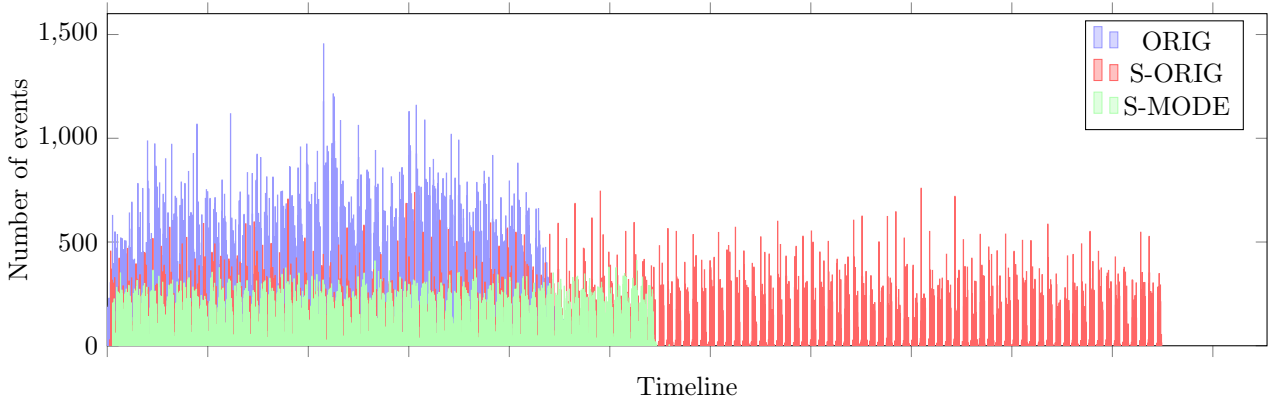

\tablename~\ref{tab:results-emd} depicts the EMD value corresponding to the start-end timestamps.
As can be seen, the repaired start times improve the EMD values of the simulated event logs (compared to simulating using the original event log) in six datasets (all datasets except MP and GOV).
All the configurations of our proposal improve the simulation with the original log in CALL, INS, and BPIC17W datasets, except for the MED-5 configuration in BPIC17W, which gets slightly worse results (0.1\% worse).
In ACR and BPIC12W, only two of our approach configurations improve the results.
The impact of the multitasking can be seen in the BNK (99\%) where only one of our approach configurations overcome the original log, and in GOV (95\%) where the best result is obtained by the simulation with the original log.
Finally, in the MP dataset, the best results are also obtained by the original log.
The meaning of a better EMD value can be better interpreted in \figurename~\ref{fig:chart-start-end-distribution}, which depicts the start-end timestamp distributions for the BPIC17W process (binned by hour).
As we can see in \figurename~\ref{fig:chart-start-end-distribution}, using the repaired event log to discover the simulation model leads to simulations reproducing better the distribution of events than using the original event log.

\tablename~\ref{tab:results-proc-time} shows the EMD values for the cycle times distribution in the event logs simulated with each technique.
Regarding this dimension, the configuration of our approach without outlier detection and using the mode as statistic (MOD) obtains the best results in all datasets but two (BNK and MP).
This shows that discovering the simulation model with the repaired start times not only allows the simulation to mimic better the original behavior in terms of event distribution, but also in therms of cycle times.

\begin{table*}[t]
    \centering
    {\footnotesize
    \caption{EMD cycle times for the simulated logs (mean of 5 simulations). The gray cells mark, for each dataset, the best result out of the repair techniques.}
    \label{tab:results-proc-time}
    \begin{tabular}{lrrrrrrrr}
              & \textbf{ACR}              & \textbf{BNK}              & \textbf{MP}                & \textbf{CALL}            & \textbf{GOV}              & \textbf{INS}              & \textbf{BPIC12W}          & \textbf{BPIC17W}          \\ \toprule \toprule
    ORIG      & 10.31                     & \cellcolor{gray!45}2.33   & \cellcolor{gray!45}15.55   & 18.19                    & 14.32                      & 10.80                     & 9.98                      & 3.95                      \\
    MED       & 8.20                      & 5.17                      & 286.27                     & 3.84                     & \cellcolor{gray!45}14.18   & 29.91                     & 95.29                     & \cellcolor{gray!45}3.93   \\
    MED-5     & 9.71                      & 5.18                      & 33.26                      & 3.88                     & 14.36                      & 11.32                     & 9.47                      & 3.96                      \\
    MED-2     & 10.50                     & 5.18                      & 18.48                      & 3.89                     & 14.36                      & 11.33                     & 10.06                     & 3.96                      \\
    MOD       & \cellcolor{gray!45}7.58   & 5.17                      & 129.49                     & \cellcolor{gray!45}3.38  & \cellcolor{gray!45}14.18   & \cellcolor{gray!45}9.78   & \cellcolor{gray!45}7.51   & \cellcolor{gray!45}3.93   \\
    MOD-5     & 9.80                      & 5.18                      & 67.27                      & 3.91                     & 14.34                      & 11.35                     & 10.06                    & 3.96                      \\
    MOD-2     & 10.52                     & 5.18                      & 21.52                      & 3.89                     & 14.35                      & 11.33                     & 10.06                    & 3.96                      \\ \bottomrule
    \end{tabular}
    }
\end{table*}

In summary, regarding the different configurations of our proposal, there is no configuration overcoming the rest in all datasets.
At an individual level, the configuration without outlier threshold and using the mode as statistic (MOD) overcomes the simulation with the original log in five of the eight datasets in terms of the timestamp distribution, and obtains the best result in six of the eight datasets in terms of cycle time.
At a generic level, in order to improve simulation results, we recommend using different configurations, as the nature of the process can affect the results.
For example, a process with highly variable activity durations and waiting times can benefit from the use of the median as a statistic, as the mode can return an unreliable value if all repaired durations have a low frequency.
Also, a low outlier threshold value reduces excessively the variability in the estimations and, as we can see in the results, this ends in less accurate simulations -- the two configurations with an outlier threshold of 2 (MED-2 and MOD-2) did not get the best result in any dataset --, although they still overcome the simulation with the original log in some cases.

\subsection{Threats to validity}

The evaluation reported above is potentially affected by the following threats to the validity.

First, regarding \textit{internal validity}, the experiments rely only on a dozen events logs. The results could be different for other datasets. To mitigate this threat, we selected logs with different sizes and characteristics and from different domains.

Second, regarding \textit{construct validity}, we used two measures of goodness based on discretized distributions. The results could be different if we employed other measures, e.g.\ measures of similarity between time series based on dynamic time warping, which provides an alternative framework for capturing rhythms in time series. 

Finally, regarding \textit{ecological validity}, the evaluation compares the simulation results against the original log. While this allows us to measure how well the simulation models replicate the as-is process, it does not allow us to assess the goodness of the simulation models in a what-if setting, i.e., predicting the performance of the process after a change.

% Let clear in the description that the baseline estimates the start time using the resource availability, but we propose a method that adds the enablement time of the activity that is being executed.
% Thus, when there is no resource information, our technique can rely on into the enablement time of the activity, while the baseline has no information to use.

% We lose: difficult to explain, better not happen, but if it does, it is maybe caused due to two causes: \textit{i)} the resource occupation is high so most of the times the resource availability is going to be the registered start time, and \textit{ii)} a wrong detection of parallelism, maybe caused by noise, or just because we are using a heuristic approach and it is not perfect, and make some cases to select the enabled time when actually there was parallelism there and we should have taken the resource availability.

    %!TEX root = main.tex

\section{Conclusion and Future Work\label{sec:conclusions}}

% Conclusions

In this paper, we presented an approach to identify the waiting time in an event log that corresponds to non-recorded processing time, and include it as processing time by repairing the start times of the activity instances.
The approach is designed to repair the start timestamps in event logs in order to capture all the processing time associated to each activity instance, with the final aim of improving the quality of automatically discovered simulation models.
%However, its scope of applicability of is wider, as it could be used to estimate the start times in an event log with only end times, supporting other process mining techniques that require both start and end timestamps~\cite{DBLP:conf/pakdd/AndrewsW17,DBLP:conf/icpm/BerkenstadtGSSW20}.

The proposal combines an approach to estimate the enablement time of activity instances in the presence of concurrency relations, and an approach to estimate the resource availability times.
In a nutshell, the underpinning assumption is that, at least for the purposes of BPS, an activity instance can be considered as started once it is enabled, and the resource who performed the activity is available.

We conducted an evaluation to analyze the impact that the start time repair has on the accuracy of automatically discovered BPS models.
To this end, we discovered a BPS model using both the original event log and the event log with repaired start times.
%We have then simulated these processes and compared the similarity between each simulated event log and the original one ---in therms of their event timestamps' and cycle times' distributions.
%In six out of the eight real-life datasets, the estimation of the start times performed by our proposal allowed the simulation to mimic the recorded process in a more faithful way.
For the majority of real-life datasets, the BPS models discovered from the event logs with repaired start times exhibited higher accuracy than the BPS models discovered from the original logs.
This observation suggests that the start times repaired by our proposal are more suitable for BPS model discovery than the start times recorded during the execution of a process in an enterprise system (subject to threats to validity discussed above).

%This proves that our approach, by estimating the start times of an event log with the premises that process simulation follows, not only allows to automatically simulate event logs with originally no start time information, but can improve the results of those event logs with already recorded activity start times.

% Future Work

%As future work, we plan to extend our proposal to scenarios where not all start times in an event log are missing (i.e.\ partially missing start times).
%use the available information in the scenarios in which the start time information is partially missing.
%In such scenarios, the recorded start times can be used to derive a probability distribution, which can then be used in the estimation of the missing start times.

A limitation of our proposal is that it requires that there is sufficient information to estimate the enablement time of each activity instance.
This is not possible for the first activity instance in a trace, since there is no previous activity instance (in the same trace) that can be used as an anchor to estimate the enablement time.
We foresee that this limitation could be addressed by applying a technique to estimate the time when each process trace arrives~\cite{DBLP:conf/bpm/MartinDC15}.
Indeed, the trace arrival time could be used as an anchor to estimate the enablement time of the first activity(ies) in the trace.

Another limitation related to the evaluation is that the BPS model automatically discovered with the original event log does not model the waiting times previous to each activity.
This could hinder its performance and increase the difference w.r.t. the BPS models discovered with the repaired start times.
For this reason, we plan to compare the improvement of our approach with a waiting time aware BPS model.
%to calculate the enablement time does not work for the first activity of each process trace, as there are no previous activity instances enabling them, hence, only the resource availability is being used to estimate their start time.
%Nevertheless, the enablement time of those activity instances actually correspond to the arrival time of the trace ---they are enabled the instant in which the process trace is created.

Furthermore, as previously acknowledged, our proposal assumes that there is no multitasking (i.e.\ resources working in more than one activity instance at a time).
The evaluation suggests that the performance of the proposed approach degrades when there is a high level of multitasking.
Addressing this shortcoming is another avenue for future work.

%It would be interesting to extend our technique to consider such behavior in order to improve the performance of the simulation.

% Calendars?

    %!TEX root = main.tex

\section*{acknowledgements}

This research is funded by the European Research Council (PIX Project).

    \bibliographystyle{plain}
    \bibliography{references}

\end{document}